# Rheology of nearly ideal 3d foams

Christopher D. Jones, Kerstin Nordstrom, and Douglas J. Durian

*Department of Physics and Astronomy, University of Pennsylvania*
*Philadelphia, PA 19104 USA*
(*October 3, 2011*)

We probe the complex rheology of nearly ideal 3d foam by flowing through a narrow column. The foams we investigate have large bubble size, to minimize the effects of coarsening, and are very dry. Foams of this type cannot be studied via conventional rheometry. The foam flows upward through a vertical rectangular column with a 4:1 cross-sectional aspect ratio, by bubbling gas through a soapy solution at the base of our apparatus. At the column's narrow surfaces are sticky boundaries, which create shear due to the zero velocity boundary condition. As expected, the flow profile between the adjacent slippery broad faces is flat, however the profile between the narrow, sticky faces exhibits a curved velocity profile that is dependent on gas flow rate. We are able to analyze a 2d velocity profile from a 3d bulk system. We employ particle image velocimetry to measure the strain rate, and compute the stress from the pressure drop along the channel, to investigate the local stress-strain relationships in a flowing foam. We find these dry foams to have a Hershel-Bulkley exponent of 0.21, which is significantly lower (more shear thinning) than other results shown in the literature for much wetter foams.

## 1. Introduction

Foam rheology has been previously investigated via simulation and experimental methodologies [1–3]. However much of the simulation work has remained completely in the realm of 2d [4–7], due to the technical difficulties of simulating fully 3-dimensional foams [8, 9]. Experimental work has also had a long history of 2d studies, with much interest in probing bubble rafts, which are considered "quasi-2d" [10–14]. Experimental 3d work has been employed via conventional rheometry in a Couette cell with very wet foams, e.g. Gillette Foamy [15, 16]. The aggregate conclusion of this body of work has shown that the Hershel-Bulkley (HB) exponent found in these studies lies within the range of 0.3 to 1 [17], however little is known about the behavior of dry 3d foams. We wish to investigate the rheological behavior of fully 3d foams as they are "ideal" foams, that is, the bubbles are polyhedra with very thin films separating them that satisfy Plateau's rules. As it is difficult to develop and maintain a very dry foam in the steady state, a foam of this type cannot be studied with conventional rheological methods. To probe the rheology of these foams, we have developed a method that addresses these issues.

## 2. Experiments

To study the rheology we have created a pressure-driven flow inside a large column. This column is a rectangular, acrylic column that is 120 cm tall (z-dimension), 10.16 cm wide (x), and 2.54 cm thick (y). The column is closed on the bottom, and is separated from another column with a divider that leaves an opening of ~5 cm at the base. In the adjacent column a gas hose is run to the bottom of the apparatus, and into the bottom of the experimental column. The bottom of the whole apparatus is filled with a solution of soapy water (2% Dawn Ultra in DI water, with 0.1% NaCl by weight; surface tension 23 dyne/cm) to a depth of ~15 cm. Nitrogen is continuously blown into the solution, controlled by a flowmeter (Cole-Parmer EW-03217-13) building a foam in the experimental column. The weight of this foam suppresses the height of the solution in the experimental column and displaces the solution in the adjacent column allowing for a simple measurement of the driving pressure.

With the creation of a fresh foam, or with a change in gas flow rate, we allow several hours of flow to ensure that a steady state has been achieved. We have independently confirmed this time is sufficient by measuring the liquid fraction over time after changes in flow rate to find the duration necessary for the difference in liquid fraction to equilibrate with drainage. Experiments were done consecutively with the same solution. Changes in the solution (as well as foam structure) were found to be negligible over the course of our investigation.

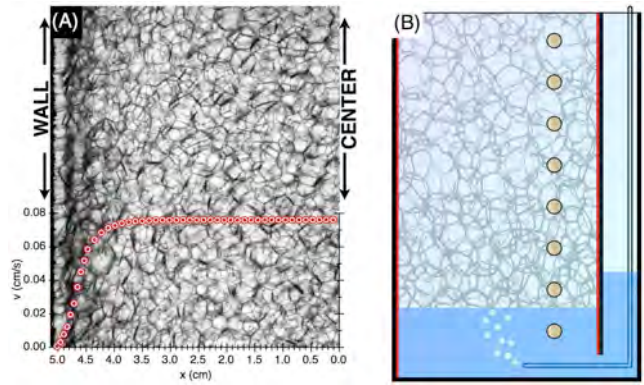

Figure 1. Snapshot of 3d foam in apparatus and schematic of apparatus (not to scale). A) Foam is driven by a nitrogen flow rate of 0.0647cm/s, with overlay of velocity profile data. The rough sandpaper can be seen on the left edge, x = 0 cm corresponds the center of the channel, and the positive z-dimension is vertical. The foam is in focus across the full thickness of the channel. The average bubble size is slightly smaller than 5 mm. B) Foam flows up a vertical channel, after being formed by gas bubbling through a soap solution at the bottom of the apparatus. The gas is introduced through a hose in an adjacent channel, which opens to the foam channel at the bottom of the apparatus. This second channel also gives us a measure of the weight of the foam based on the displacement of solution in this channel. Red lines denote sticky surfaces which create a no-slip condition at those boundaries, and brass electrodes are evenly spaced up the vertical z-dimension.

The narrow sides of the column are covered with ceramic aluminum oxide sandpaper belt of 36 grit (~0.5 mm particle diameter), which gives a reliable no-slip condition. The wide sides are left as bare acrylic which result in a very slippery surface. It is important to note that the sandpaper does not tear the soap films or pop the bubbles.

A series of brass electrodes, 0.64 cm in diameter, are used to infer the liquid fraction via measurement of electrical conductivity. The conductivity of both solution and foam are measured with a 1715 LCR Digibridge impedance meter from QuadTech at 1.0 V and 1 kHz frequency [18]. We use the measured conductivities to calculate the liquid fraction, via Lemlich's approximation, $\frac{\sigma_f}{\sigma_l} = \frac{1}{3}\varphi_l$, where $\sigma_f$ and $\sigma_l$ are the foam and liquid conductivities, and the volume liquid fraction is $\varphi_l$[19].

The column is backlit by diffuse white fluorescent light, and a Vision Research Phantom v9.0 camera with a Nikkor AF-S 55-200 mm lens was used to take video of the flowing foam at 4 frames per second. The region of interest used for video acquisition was at a height found to have negligible change in liquid fraction for heights around the region, with a total size of 1100 by 768 pixels (roughly 8 by 5.5 cm).

To investigate the foam's structure, we calculate the autocorrelation function of the foam images in the direction of foam flow (z-dimension). This analysis was done in strips, 1.2 mm (16 pixels) wide, and the full length of the image in z, in order to show any possible changes in foam structure across the cell. Several videos of foam were taken, and averages of the autocorrelation were taken as the foam had flowed approximately half of the video frame. We averaged over approximately 1,000 unique foams to obtain the smooth autocorrelation function shown in figure 2. The first peak in the autocorrelation tells us the average bubble size of the foam, and the initial decay of the function relates to the thickness of the Plateau Borders.

To understand the flow properties of the foam, we measure the velocity profile across the cell. A tracking method cannot be used here because the Plateau Borders (PBs) are irregular and interconnected. The relative spacing of the PBs does not change appreciably from one frame to the next, so it stands to reason that we can use a correlation technique to extract how far a foam has traveled in one frame. We use this technique, a form of Particle Image Velocimetry (PIV) using a homebuilt code written in the LabVIEW (National Instruments) programming environment[20, 21]. To extract the velocity profile, the movie is broken into constituent frames. Each frame is cut into strips of varying width and each strip encompasses the entire z-dimension of the frame, and a specific, strip location in the x-direction. We are interested in the displacements of features in a strip from one frame to the next. We denote each strip by $I_s(f)$ where $I$ is a matrix representing the 8-bit grayscale intensity levels in a strip $s$ at frame number $f$. The displacement between $I_s(f)$ and $I_s(f + \Delta f)$, using cross-correlation. Calculation of the cross-correlation, $C_s$, of $I_s(f)$ and $I_s(f + \Delta f)$ is straightforward, using a 2d spatial Fourier transform F:

$$C_s(f, f + \Delta f, x, y) = F^{-1}\{F[I(f, x, y)]^* \cdot F[I(f + \Delta f, x, y)]\}.$$

$C_s$ is a peaked function where the $(x, y)$ position of the maximum gives the relative displacement between the two strips. This maximum is found to subpixel accuracy via a parabolic fit. We repeat this on one strip for all frames and take the average to give the velocity at that x-position. Repeating this for different strips fills in the rest of the velocity profile.

The width of the strips is adapted to the strain rate $\dot{\gamma}$; near the center of the cell where $\dot{\gamma}$ is small, wider strips are used. Overlapping of strips allows for a consistent resolution in x from center to edge of cell. This results in getting velocities for every x position across the cell, with a displacement near the center of ~1.2 mm.

To improve the accuracy of our velocity measurement we vary the time interval $\Delta f$, or delay time, between cross correlation frames. In the range of delay times where the ratio of calculated displacement over step size is a constant, we fit our data to a constant which we take as the speed, and the uncertainty of the fit is taken as our error. By doing this we are able to insure that a large enough time step has been used to optimize the PIV analysis, and to confirm that the speed of that strip is consistently measured over a broad range of time steps.

The results are shown in figure 3a where the velocity profile across the cell width is plotted versus x. The velocity profile is typical of plug flow. While the resultant velocity profile is quite smooth, there are not enough points to make finite differencing useful to calculate the first derivative. However, since it is smooth, we can use a polynomial expansion to get the derivatives. We have automated the subsequent process in LabVIEW as well. Each point on the profile is assigned a fitting window over several points. These points are fit with a weighted linear least squares fit to a 3rd degree polynomial. The points are weighted by both relative uncertainties from the previous fit to a constant speed ($W = 1/\sigma^2$) and by proximity to the central point. The further away from the central point, the less influence it has in the fit via a Gaussian drop-off in weights. Ultimately, this fitting routine gives values and uncertainties for the first derivative (the strain rate) at each point [22]. The resulting strain rate versus x curves are shown in figure 3b.

## 3. Results and analysis

At higher foam flow rates, the foam will be wetter. This is due to the foam carrying liquid with it as it flows, and as the flow rate increases, the drainage has less of an effect on the overall wetness of the foam. So, a slower flow rate will result in a dryer foam at points high in the channel, while a faster flow rate will give a wetter foam. For this reason the liquid fractions of the two flow rates we show here are different. Our lower flow rate is 0.0647 cm/s, the average liquid fraction $<\varepsilon>$ is 1.07e-3, and liquid fraction at the region of interest is 3.01e-4. The higher flow rate is 0.115 cm/s, $<\varepsilon> = 1.35$e-3, and at the region of interest $\varepsilon = 5.36$e-4.

The average autocorrelation function for all strips in x across the cell is shown in figure 2. The inset shows the position of the first peak in the autocorrelation function with respect to x-position. Image autocorrelation analysis was done at varying heights along the length of the channel to confirm that the fundamental structure of the foam does not change with height (results not shown). Once this confirmation was completed, a height was chosen where the liquid fraction as a function of height curve was relatively flat, at a height approximately 45 cm above the solution-foam interface. The foams have very similar average bubble sizes, at 0.39 cm for the slow flow, and 0.42 cm for the faster flow rate. Even though the two foams are at different gas flow rates, and therefore differing liquid fraction, the structures are very similar, as well as being consistent in x. In this way we verify that the the fundamental structure of the foam is not different at different locations in our column, or at different flow rates, even though the foam is, in fact, not identical otherwise.

Fortunately, differences in foam wetness do not have a significant impact on the structure of the foam, as in both cases the foam is still very dry, and the resultant rheological behavior does not significantly deviate. With the consistency in structure confirmed, we are able to start analyzing the velocity profile shown in figure 3. The velocity profile for the two flow rates we show are smooth, and with the application of our polynomial fitting program it is possible to see what minimum strain rates we are able to investigate before the strain rate becomes ill-defined. This minimum possible strain rate is quite far into the plug, and is within just a few bubbles from the centerline of our column.

With the strain rate, and positions now determined, we must find the stress imposed by the zero-slip boundary at the edge of the column. A simple force balance

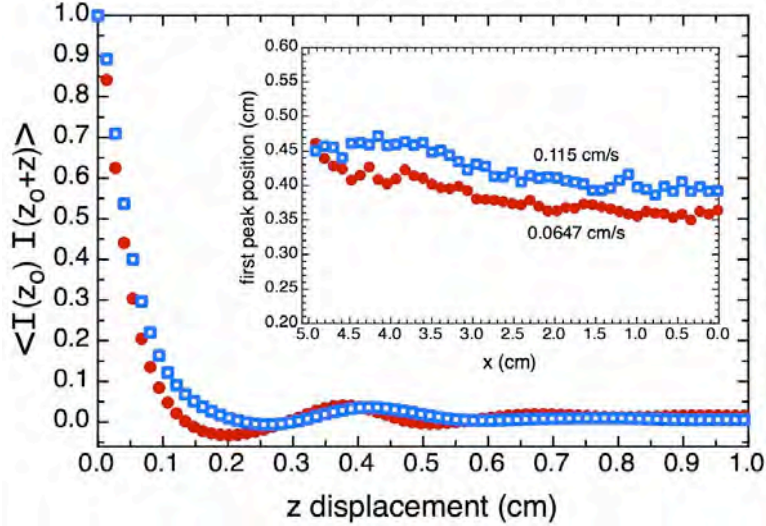

Figure 2. Plot of the normalized image autocorrelation function showing initial decay and first peak, where the horizontal axis indicates displacement along the z dimension, or the direction of foam flow. Inset shows position of first peak across x-dimension of cell. Red circles (blue squares) denote a gas flow rate of 0.0647 cm/s (0.115 cm/s). Plots show that both flow rates have very similar foam structure, and the first peak position gives the average bubble size, which is also consistent across the width of the channel. Averaging the bubble size across x gives an average bubble size of 3.9 mm (red circles) and 4.2 mm (blue squares).

analysis, of the upward force driving the foam, from -x to x, was performed:

$$\Sigma F = 0 = (\rho g h)(2xw) - (<\varepsilon> \rho g \cdot 2xwH) - 2\sigma w H - 2\sigma_f \cdot 2xH.$$

Where the solution density is $\rho$ (1 g/ml), the suppression height from the weight of the foam displacing the solution at the bottom of the column h, the distance from the centerline of the column x, the width of the column w, the total height of the foam in the column H, and the drag along the two broad slippery faces $\sigma_f$. This reduces to give us the stress, as a function of x:

$$\sigma = \rho g (\frac{h}{H} - <\varepsilon>)x - \sigma_f \cdot \frac{2x}{w}.$$

We expect the face drag to be approximately $(\gamma/R)(\eta v/\gamma)^{2/3}$ where $\gamma$ and $\eta$ are the surface tension and viscosity of the solution, respectively, and R is the bubble radius. The ratio of the face drag term to the first term in our stress solution, above, then comes out to approximately 0.005. With this we determine the face drag along our slippery faces to be a negligible addition to the total stress.

The stress versus strain rate relationship of these foam flows is shown in figure 4. Even though the liquid fraction of the two flow rates is different, the fundamental structure in both is very similar, and hence both data sets collapse. There are still small differences, especially in the value of the yield stress between the two. We fit the data, ignoring the portion in both sets where the strain rate turns back on itself at its highest values, to a Herschel-Bulkley (HB) relation, $\sigma = \sigma_y + c_v \dot{\gamma}^a$, where $\sigma_y$ is yield stress, $c_v$ is the consistency, and $a$ is the HB exponent. Our fit results give us $a$=0.213 ±0.05, which is close to the minimum experimental value measured in previous literature, and is far below the theoretical prediction of Schwartz and Princen [23], as well as the experimental observations of Princen and Kiss [24]. Our measured value is even further below a value used frequently, of 1, when the studied foams exhibit a Bingham response [16].

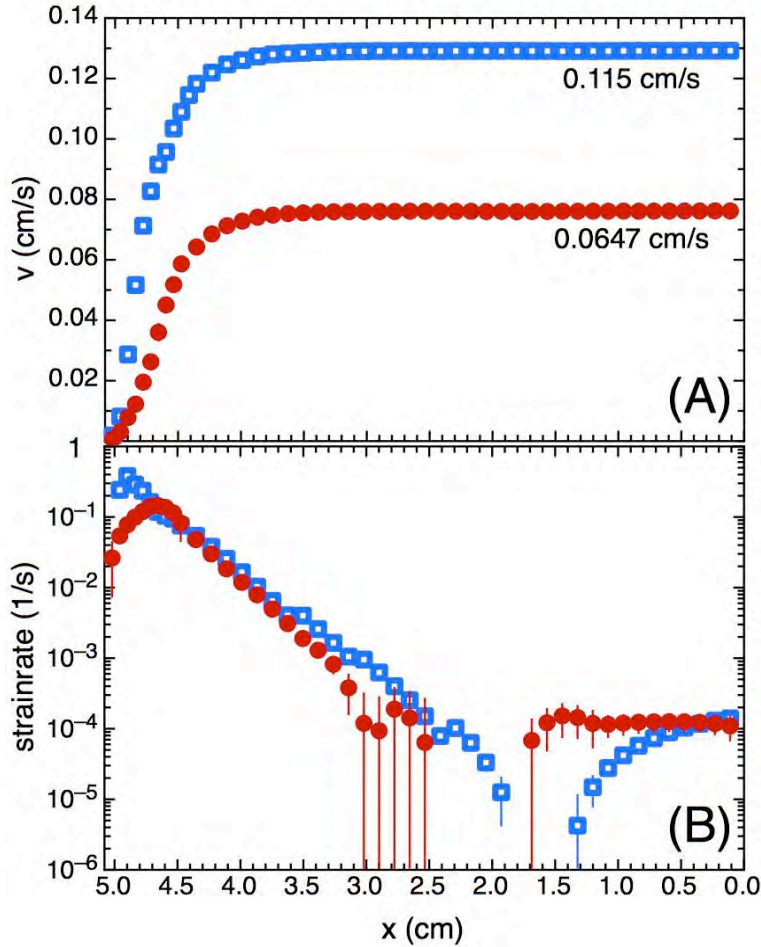

Figure 3. Plots of the (a) velocity and (b) strainrate profiles for both foam flow rates from the sticky boundary to the center of the cell. (a) Velocity profiles obtained from PIV analysis. When normalized the two velocity profiles collapse, giving the same plug dimensions. Using our PolyFit program we obtain the (b) strainrate profile from the velocity profile. Well inside the plug the strainrate becomes ill-defined, and we are unable to obtain reliable strainrate numbers at the point where the strainrate changes sign. However, we can use strainrates at 2.5 cm (2.0 cm) from the center of the channel for the slower (faster) gas flow rate. The slower flow rate (red circles) shows a turning down of strainrate on approaching the cell wall at x = 5.08 cm. This is a result of the fact that the cell boundary has a monolayer of bubbles constrained to it, while the bulk of the cell has a random collection of bubble positions. This first layer of bubbles has a different rheological response from the bulk. This can also be seen slightly in the higher flow rate (blue squares), however at higher flow rates the bubbles attached to the boundary tend to be somewhat smaller, diminishing both the region of different (more highly aligned) structure, and reducing the size of the films responding to shear.

An HB exponent of 0.24 was measured by Denkov, et al. [17] for foams with immobile surfaces, which is explained by the dependence on surface mobility coupled with viscous friction inside the bubbles. In the case of a surface slip condition, they observe an exponent of 0.42, far closer to the conventional expectation [10]. It is clear that surface interactions and liquid fraction feature strongly into the measured exponent [25, 26]. Our low result indicates the significance of a true stick condition at the surface, and the contribution of extremely low liquid fractions. While exponents of 0.4 to 1 are readily used as approximations, we argue that when approximating ideal foams a more realistic number of 0.2 should be used, based on our empirical result.

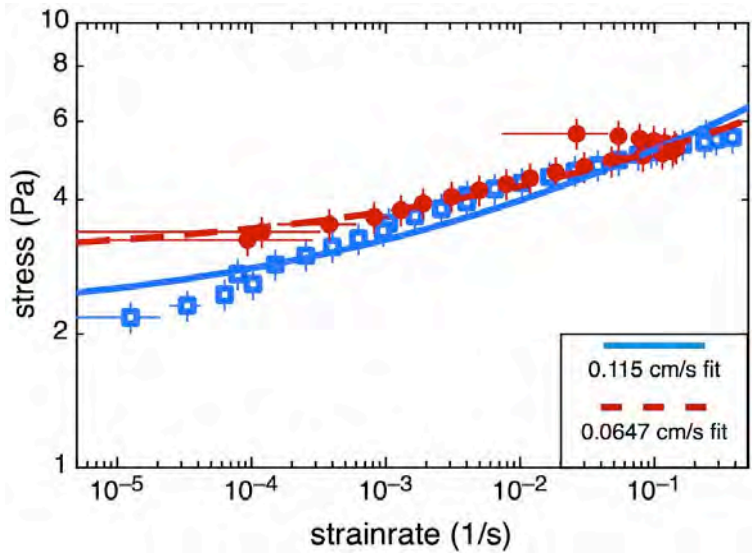

Figure 4. Plots of stress versus strainrate for both flows, along with fits to a Hershel-Bulkley relation. In both cases the H-B exponent is found to be 0.213+/-0.05. Gas flow rate of 0.0647 cm/s corresponds to the red circles, with the dotted fit curve, and the fitted yield stress is 2.94+/-0.22 Pa. The gas flow rate of 0.115 cm/s corresponds to the blue squares, with the solid fit curve, and the fitted yield stress is 2.10+/-0.09.

## 4. Conclusion

We have developed an apparatus that allows us to investigate 3d foams that are drier, and therefore more ideal, than foams that have been studied by other, more traditional techniques. This does limit us to larger bubble sizes, in order to minimize the effects of coarsening on experimental time scales, however, this is a system that we expect to have much future use because of the lack of limitations this technique has compared to other probes. While it is still not yet clear how exactly the HB exponent does depend on specific conditions, we can verify that surface mobility and foam wetness contribute significantly with these very dry 3d foams.

## 5. Acknowledgements

Work supported by NASA.